# Data Report: Permeability, Compressibility, and Friction Coefficient Measurements under Confining Pressure and Strain, Leg 190, Nankai Trough[1]

Sylvain Bourlange,[2, 3] Laurence Jouniaux,[2] and Pierre Henry[2, 4]

## ABSTRACT

Permeability measured on three samples in a triaxial cell under effective confining pressure from 0.2 to 2.5 MPa ranges from $10^{-18}$ to $10^{-19}$ $m^2$. Overall, results indicate that permeability decreases with effective confining pressure up to 1.5 MPa; however, measurements at low effective pressure are too dispersed to yield a precise general relationship between permeability and pressure. When the effective pressure is increased from 1.5 to 2.5 MPa, permeability is roughly constant ($\sim$1–4 $\times$ $10^{-19}$ $m^2$). Samples deformed in the triaxial cell developed slickenlined fractures, and permeability measurements were performed before and after failure. A permeability increase is observed when the sample fails under low effective confining pressure (0.2 MPa), but not under effective pressure corresponding to the overburden stress. Under isotropic stress conditions, permeability decrease related to fracture closure occurs at a relatively high effective pressure of $\sim$1.5 MPa. Coefficients of friction on the fractures formed in the triaxial cell are $\sim$0.4.

## INTRODUCTION

Permeability and storage coefficients are essential parameters controlling fluid flow and pore pressure regime. Knowledge of these param-eters under in situ conditions is therefore important in understanding deformation processes in accretionary complexes. Most permeability measurements on samples from the Nankai accretionary complex have been performed without pressure confinement (Taylor and Fischer, 1993) or at low confining pressure (<1 MPa) (Gamage and Screaton, this volume; Karig, 1993). Measurements of permeability under 1–5 MPa effective confining stress give lower values (Byrne et al., 1993). Here, we report measurements performed in the 0.5- to 2.5-MPa range in a triaxial cell with the main purpose of approaching in situ stress conditions. These measurements are indicative of the permeability of the unfractured sediment and may, therefore, constrain fluid flow out of the fault zones through the wallrock. The permeability of fractured zones may be higher, especially if the fractures are dilated because of a high fluid pressure. To investigate the effect of fracturing, samples were failed along drained paths at low (0.2 MPa) and relatively high (2.5 MPa) effective pressure and permeability was measured before and after fracturation. Slickenlined fractures formed during the failure tests with visual aspects similar to fractures observed on core samples, and it was possible to determine the friction coefficient of these surfaces. Experimental difficulties and time constraints limited the number of samples that could be processed; therefore, interpretations are preliminary.

[2]Laboratoire de Geologie, Ecole Normale Superieure UMR 8538, 24 rue Lhomond, 75231 Paris Cedex 05, France. Correspondence author: bourlang@geologie.ens.fr
[3]Present address: LIAD–ENSG, Rue du doyen Marcel Roubault, BP 40, 54501 Vandoeuvre, France.
[4]Present address: Chaire de Geodynamique, College de France, Europole de l'Arbois, BP 80, 13545 Aix en Provence Cedex 04, France.

## METHODS AND SAMPLES

Permeabilities in this study were measured using a pulse decay method (Brace et al., 1968; Jouniaux et al., 1994, 1995). The pulse is a small step change (0.1–0.3 MPa) of differential fluid pressure imposed between pressure vessels connected at the ends of the sample. When a pressure pulse $\Delta P_0$ is applied, the differential pressure $\Delta P(t)$ decays exponentially as a function of time, $t$:

$$\Delta P(t) = 2\Delta P_0 V_2/(V_1 + V_2)\, e^{-mt},$$

where

$V_1$, $V_2$ = the upstream and downstream reservoir volumes ($V_1 = V_2 = 50 \times 10^{-6}$ m$^3$ in our experimental setup),
$t$     = time, and
m    = a decay time constant (Fig. F1).

Plotting the decay curve in terms of ln[$\Delta P(t)$] vs. time $t$ yields a straight line having a slope m, and the permeability $k$ can be determined by

$$k = m\mu(L/A) \times [\mathrm{Cup} \times \mathrm{Cd}/(\mathrm{Cup}+\mathrm{Cd})],$$

where

$L$  = length of the sample,
$A$  = cross-sectional area of the sample,
m = dynamic viscosity of pore fluid at temperature measurement (10$^{-3}$ Pa·s at 20°C), and
[Cup $\times$ Cd/(Cup+Cd)] = storage of the pressure vessels (2.4 $\times$ 10$^{-14}$ m$^3$/Pa).

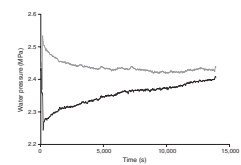

F1. Permeability measurement, p. 9.

Two GDS pressure controllers (50 cm³ internal volume) in standby mode were used as constant-volume pressure vessels during pulse decay measurements. Forty-one measurements were performed with the pulse decay method at various levels of effective confining pressure (confining pressure – pore pressure) from 0.34 to 2.4 MPa.

We discarded measurements when an exponential could not be fitted to the data. The largest deviations from the ideal exponential curve correlate with room temperature variations and are attributed to the thermal expansion of water within the pressure controllers (or of the pressure controllers themselves). The pulse decay method requires a small pulse initial pressure difference (10%) compared to the effective pressure. Because the pressure pulse is at least 0.1 MPa (limited by the precision of the pressure gauges), measurements performed at effective confining pressure <1 MPa do not follow this condition. These measurements were retained when the exponential fit was good but yielded scattered results (Fig. F2).

For comparison, two permeability measurements were performed from steady-state flow under a constant pressure gradient ΔP (0.1–0.28 MPa) with GDS pressure controllers in locked pressure mode. Time needed for these measurements is about twice the time needed using the pulse decay method. Permeability was calculated using Darcy's law:

$$k = Q\mu L/(A \times \Delta P).$$

Measurements were performed on three samples (diameter = 25 mm) from Ocean Drilling Program (ODP) Leg 190 (Moore, Taira, Klaus, et al., 2001; Moore et al., 2001) cored either in the vertical direction (v) or in the horizontal direction (h) (Table T1). The overburden stress was calculated by integrating the bulk density with depth. The effective overburden stress range is estimated assuming an overpressure ratio between 0 and 0.42 (maximum overpressure ratio determined for the underthrust sediments by Screaton et al. [2002]).

# RESULTS

## Permeability

Permeability measurements were performed on sample 190-1174B-42R-1, 93–95 cm (vertical), under increasing isotropic pressure from effective pressures of 0.5–1.5 MPa over a period of 10 days (Fig. F2A). This sample was maintained at 0.5 MPa effective pressure for 8 days and absorbed 1262 mm³ water without significant total volume change, showing that the sample was not fully saturated at the beginning of the experiment. We calculated an initial water saturation of 86%. Measurements made on Sample 190-1174B-42R-1, 93–95 cm (vertical), allowed a comparison of steady-state flow and pressure decay methods. The two measurements performed using the flow method (1.2 × 10⁻¹⁸ m² at an effective pressure of 0.6 MPa and 4.1 × 10⁻¹⁹ m² at an effective pressure of 1.41 MPa) are at the upper limit of the range of measurements done using the pulse decay method (3.6–4.5 × 10⁻¹⁹ m² at 0.5 MPa effective pressure and 1.6–3.7 × 10⁻¹⁹ m² at 1.5 MPa effective pressure) but are compatible. The 1.2 ± 0.2 × 10⁻¹⁸ m² permeability value at 0.6 MPa effective pressure we obtained using the flow method is close to the 2.3 × 10⁻¹⁸ m² permeability measured using the same method on Sample 190-

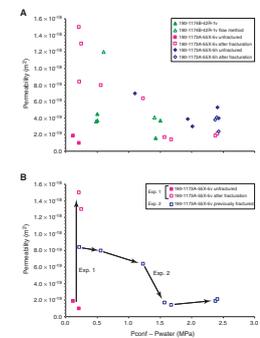

1174B-42R-3, 133–150 cm (538 meters below seafloor [mbsf]), at the same effective pressure (**Gamage and Screaton,** this volume).

Sample 190-1173A-55X-5, 135–150 cm (vertical), was fractured during a failure test (experiment 1) performed at a low confining effective stress (0.2 MPa). This fracturation induced a single order of magnitude permeability increase from 1–2 × 10⁻¹⁹ m² to ~1.4 × 10⁻¹⁸ m² (Fig. **F2B,** exp. 1). The sample was removed from the triaxial press to observe the fracture (Fig. **F3**). This same sample (called "previously fractured" in Table **T2**) was used 2 days later for new permeability measurements under increasing isotropic conditions from 0.2 to 2.4 MPa effective pressure for 6 days (experiment 2). The evolution of confining and pore pressures and the permeability measurements are shown in Figure **F4.** Permeability decreased by nearly one order of magnitude from ~10⁻¹⁸ to ~10⁻¹⁹ m² from effective pressures of 0.5 to 1.5 MPa (Fig. **F2B,** exp. 2). Then the permeability remained nearly constant (~1.8 × 10⁻¹⁹ m²) with increasing effective pressure.

Sample 190-1173A-55X-5, 135–150 cm (horizontal), was loaded along an isotropic path to a high confining effective stress (2.5 MPa). Permeability was measured first under isotropic conditions and then under deviatoric stress before and after failure. Failure occurred at 5.0 MPa of effective axial stress and 1.5% of deformation. The deformation rate was 4.2 × 10⁻⁹ s⁻¹ during the hydrostatic stage (14 days) and 1.9–3.2 × 10⁻⁸ s⁻¹ during the deviatoric stage (16 days). Permeability decreased by a factor of two (from ~8 to ~4 × 10⁻¹⁹ m²) during isotropic loading as effective pressure increased from 1 to 2 MPa and then stayed roughly constant. Permeability decreased during failure and then recovered as the sample was deformed beyond the failure point. Photographs showing the fractures were taken after the experiments (Fig. **F5**).

All our permeability measurements performed in the triaxial cell are lower than measurements performed at a low effective stress (Taylor and Fisher, 1993; **Gamage and Screaton,** this volume), which range from 10⁻¹⁶ to 10⁻¹⁸ m² for samples from ~600 mbsf. However, they are in the same range as values measured at <1–5 MPa effective confining stress (Byrne et al., 1993), ranging 1.3 × 10⁻²⁰ to 1.3 × 10⁻¹⁸ m². Byrne et al. (1993) also observed transient permeability variations with strain.

Overall results indicate permeability decreases with effective confining pressure as high as 1.5 MPa (Fig. **F2A**). This permeability reduction occurs with cumulative void decreases of 0.015–0.054 depending on the sample (Table **T2**). However, measurements at low effective pressure are too dispersed to yield a precise general relationship between pressure and permeability. When the effective pressure is increased as high as 2.5 Mpa, permeability is roughly constant (1–4 × 10⁻¹⁹ m²) but the void ratio continues to decrease, reaching a final cumulative void ratio decrease of 0.08–0.10. Our results also suggest that fracturing does not affect permeability when the effective stress is more than ~1.5 MPa. They also show the enhancement of permeability by fracturing when effective stress is low, followed by the permeability decrease with the increase in effective stress, interpreted as fracture closure.

## Bulk Compressibility Measurements

The bulk compressibilities of the samples were estimated using the volume of pore fluid expelled from the sample ($\Delta V$) in response to an increase of the effective confining pressure in isotropic stress conditions and before the yield point. The compressibility is calculated as

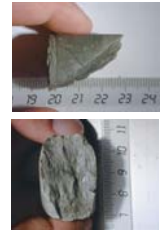

**F3.** Sample 190-1173A-55X5, 135–150 cm (horizontal), p. 11.

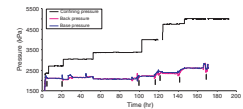

**F4.** Evolution of confining and pore pressures, p. 12.

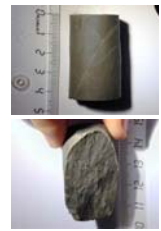

**F5.** Sample 190-1173A-55X5, 135–150 cm (vertical), p. 13.

$$\alpha = 1/V \times (\Delta V/\Delta P),$$

where $V$ = the volume of the sample. Results show a value of $1.1 \pm 0.2 \times 10^{-8}$ Pa$^{-1}$ for Sample 190-1174B-42R-1, 93–95 cm; $1.5 \pm 0.2 \times 10^{-8}$ MPa$^{-1}$ for Sample 190-1173A-55X-5, 135–150 cm, in the vertical direction; and $1.4 \pm 0.5 \times 10^{-8}$ MPa$^{-1}$ for the same sample in the horizontal direction. These compressibility values correspond to the elastic deformation of the sample. Compressibilities determined on the unloading curve were in the same range, confirming that the compressibilities were measured in the elastic domain of the samples.

These estimations can be compared with values deduced from the stiffness modulus derived from consolidation tests realized after Leg 131 (Moran et al., 1993). The compressibility is the inverse of the stiffness modulus. Before the yield point, Sample 131-808C-23R-3 (514 mbsf) showed a stiffness modulus of 50–250 MPa, corresponding to a compressibility of $4.0 \times 10^{-9}$ to $2.0 \times 10^{-8}$ Pa$^{-1}$. These different estimations remain in the same range.

## Friction Coefficients

Friction coefficients were calculated for vertically and horizontally oriented Samples 190-1173A-55X-5, 135–150 cm (horizontal), and the same sample in the vertical direction. These samples were subjected to drained axial compression in the triaxial cell and developed slickenlined faultlike fractures during failure (Figs. F3, F5). The shear and normal stresses resolved on a rupture plane inclined at an angle of $\alpha$ from the vertical are as follows (Jaeger, 1959):

$$\tau = [(\sigma_1 - \sigma_3)/2] \times \sin(2\alpha), \text{ and}$$

$$\sigma_n = \{[(\sigma_1 + \sigma_3)/2] - [(\sigma_1 - \sigma_3)/2]\} \times \cos(2\alpha),$$

where

$\sigma_1$ = effective axial stress (axial stress – pore pressure) and
$\sigma_3$ = effective confining pressure.

As axial strain increases, axial stress reaches a peak stress corresponding to the failure of the sample. Then the stress reaches a plateau lower than the peak stress, corresponding to an axial strain increase at constant stresses. This plateau is interpreted as steady-state sliding on the previously formed fracture. The friction coefficient, $\mu$, is calculated at this plateau (Byerlee, 1978):

$$\mu = \tau/\sigma_n.$$

Three friction coefficients were calculated on two samples (Table T3). Sample 190-1173A-55X-5 (horizontal) presented a friction coefficient of 0.40. The previously fractured Sample 190-1173A-55X-5 (vertical) presented a strain-stress curve with a plateau at 4.9 MPa axial effective stress, interrupted by a 45% drop in axial stress, followed by a new axial stress buildup to a second plateau at 5.2 MPa. The sudden drop in axial stress could be the result of oversliding (unstable sliding) on the fracture. The friction coefficients calculated at these two plateaus are 0.37 and 0.40. These friction coefficients are consistent with measurements per-formed (usually at much higher effective pressure on the order of 100 MPa) on pure Ca smectite gouge during velocity-stepping direct shear experiments (Saffer et al., 2001) or on a gouge of a saturated mixture of ~60% quartz and ~40% montmorillonite (Logan and Ranenzahn, 1987) during steady-state sliding and on saturated illite (Morrow et al., 1992). Kopf and Brown (2003) measured friction coefficients of ~0.14 for purified smectite and ~0.25 for illite under a range of effective pressures of 1–30 MPa. They estimated, based on ring shear experiments and mineralogical composition, the friction coefficient on Nankai décollement to be 0.30–0.32. These values, lower than our measurements, were performed using elevated displacement. The values of Kopf and Brown (2003) may be representative of well-developed faults like the décollement, whereas our measured friction coefficients may be representative of fractures with shorter sliding, and therefore closer to initial failure conditions.

# CONCLUSION

Permeability measurements show values of ~2–6 × 10$^{-19}$ m² when the effective pressure is ~2.5 MPa and the drained compressibility of the sample is ~1.3 × 10$^{-8}$ Pa$^{-1}$. If these values are representative of permeability and compressibility of fault wallrock under in situ conditions, diffusion of pore pressure in the wallrock would affect a characteristic length of 20–50 m in ~100–1000 yr. Considering that the fault zone itself is 20–30 m wide and that the diffusion distance varies with the square root of time, equilibration of pore pressure between the fault zone should be achieved in <100,000 yr. This implies that a higher pore pressure in the fault zone than in the surrounding sediments can be maintained only during transient events. These transient events may relate to slip events and to décollement propagation episodes (Bourlange et al., 2003).

Failure tests show that permeability increases as a result of the formation of a faultlike slip plane in the sample when confining stress is low. Sample 190-1173A-55X-5 (vertical) was fractured at 0.2-MPa confining effective stress, and this induced an increase of permeability by one order of magnitude. This fractured sample was then used in the triaxial cell for permeability measurements under increasing confining pressure. During this second phase, permeability decreased sharply between 1.23 and 1.57 MPa, whereas the sample experienced only a small (0.01) void ratio change (see Table T2; Fig. F2). This sharp permeability decrease likely corresponds to fracture closure. This suggests that once hydraulically conductive fractures have been formed, they may remain hydraulically conductive up to a relatively high confining stress (1.2 MPa). This permeability cycle presents similarity with transient increases in expelled water flow observed during sample loading (Byrne et al., 1993). These observations were interpreted as a transient increase in permeability associated with dilation and shear zone formation in the sample.

Friction coefficients determined in this study (0.37–0.40) are in the range expected for the clay-rich lithology and are not exceptionally low. This suggests a high fluid pressure is required inside the fault zone for decoupling at the décollement level. If a Coulomb wedge model is used (Dahlen, 1984), sliding along the décollement is allowed only when the excess pore pressure ratio, $\lambda_b*$, at the décollement is >0.6:

$$\lambda_b* = (Pf - Ph)/(Pt - Ph),$$

where

Pf = interstitial fluid pressure,
Ph = hydrostatic fluid pressure, and
Pt = lithostatic pressure.

This value is significantly higher than the 0.47 value estimated from compaction curves (Screaton et al., 2002) at ODP Site 808. This suggests a higher pore pressure in the fault zone during episodes of sliding.

## ACKNOWLEDGMENTS

We wish to thank Guy Marolleau at École Normale Supérieure (ENS) for maintaining our suffering triaxial cell, Jean Philippe Avouac and Frederic Perrier at the Commissariat à l'Ènergie Atomique/Direction des Applications Militaires (CEA/DAM) who provided us with three GDS cell controllers. This study was partly funded by Centre Nationale de la Recherche Scientifique (CNRS) and by the Geosciences "ad hoc" committee of Institut National des Sciences de l'Univers (INSU). This research used samples and data provided by the Ocean Drilling Program (ODP). The ODP is sponsored by the U.S. National Science Foundation (NSF) and participating countries under management of Joint Oceanographic Institutions (JOI), Inc.

**Figure F1.** Example of permeability measurement for the previously fractured Sample 190-1173A-55X-5, 135–150 cm (vertical), using the pulse decay method at 1.57 MPa effective pressure and 1.6 MPa effective axial stress. The two curves correspond to the pore pressure at both ends of the sample. The permeability deduced is $1.7 \times 10^{-19}$ m².

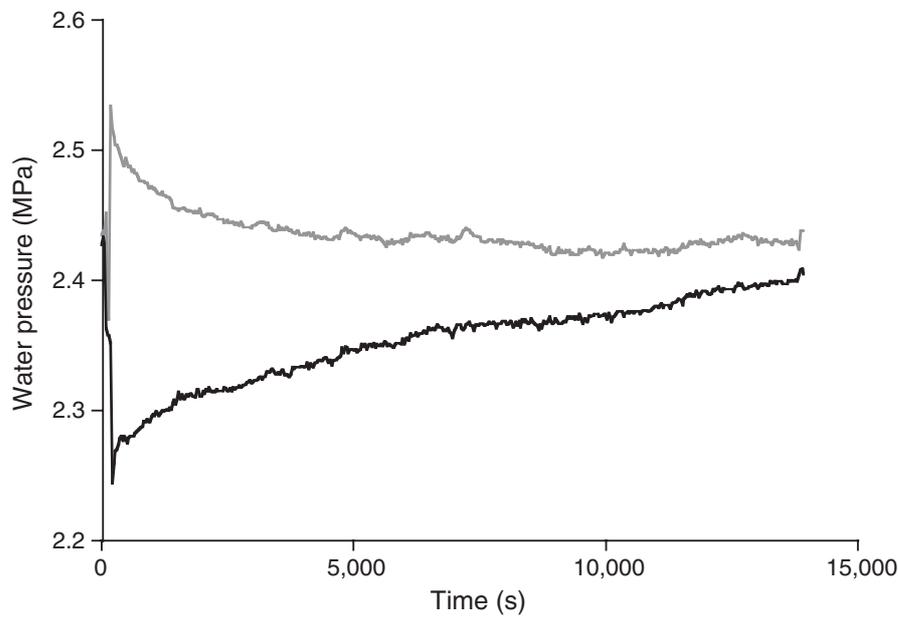

**Figure F2. A.** Permeability values as a function of the effective pressure for the four experiments. **B.** Permeability and effective pressure paths for experiments 1 and 2 on Sample 190-1173A-55X-5, 135–150 cm (vertical). v = vertical, h = horizontal. Pconf = confining pressure, P water = pore pressure.

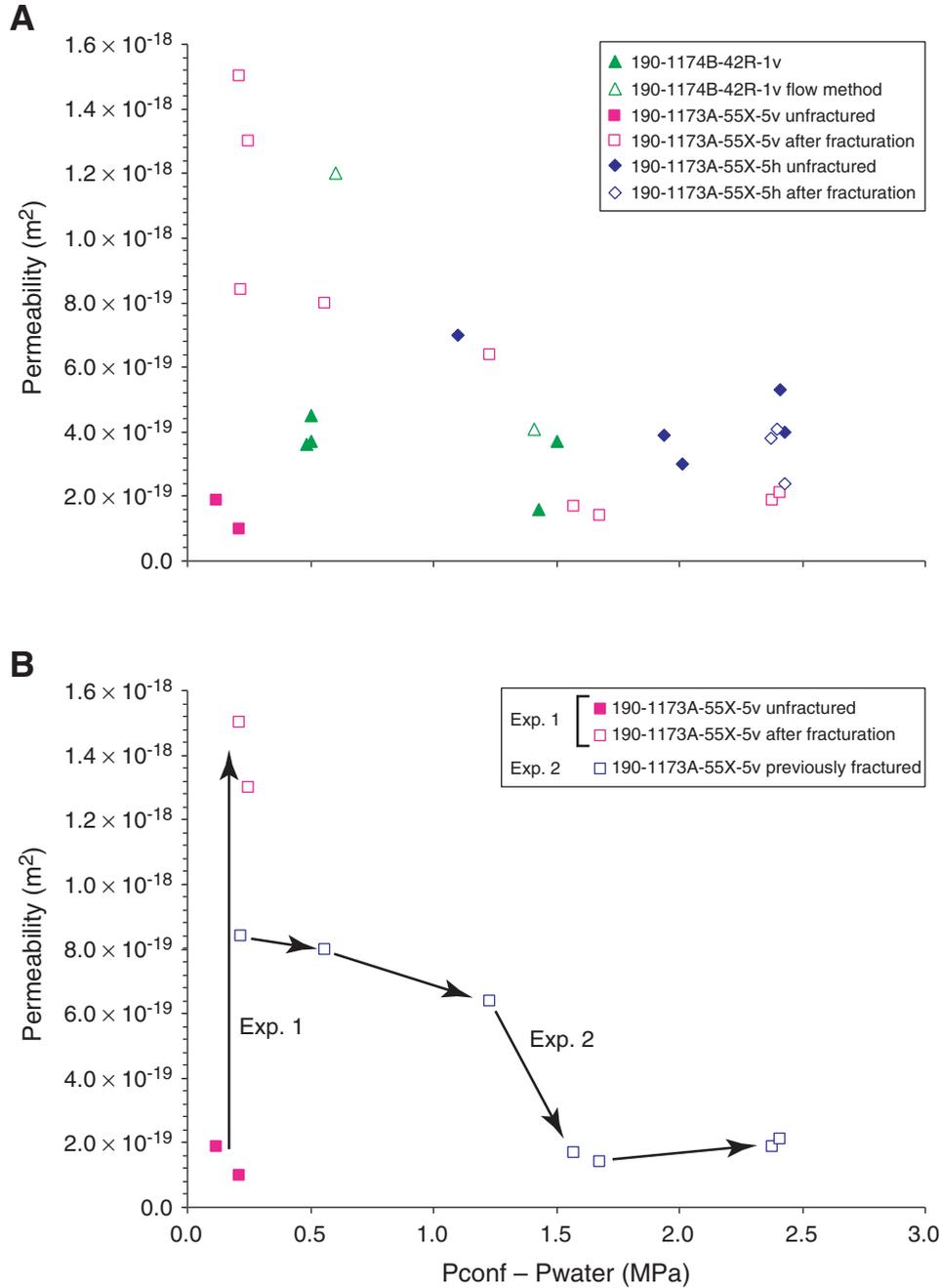

**Figure F3.** Sample 190-1173A-55X5, 135–150 cm (horizontal), after the experiment. Upper photograph shows the main fracture plane with a 30° angle, and two auxiliary fractures (one parallel to the main fracture and one conjugate at the bottom left of the sample). The lower photograph show the rupture plane, with visible striation.

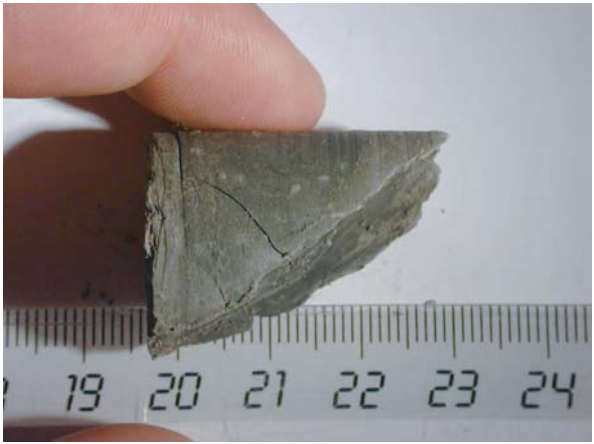

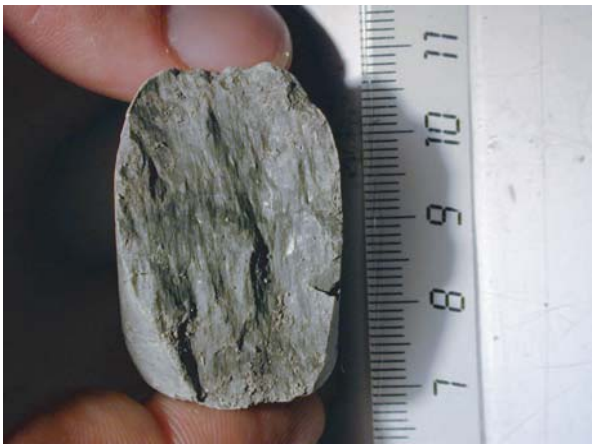

**Figure F4.** Evolution of confining and pore pressures during experiment 2 on Sample 190-1173A-55X-5, 135–150 cm (vertical; previously fractured). Permeability measurements are indicated by arrows. The corresponding effective pressures of these measurements are, respectively, 0.22, 0.56, 1.23, 1.68, 1.57, 2.4, and 2.4 MPa.

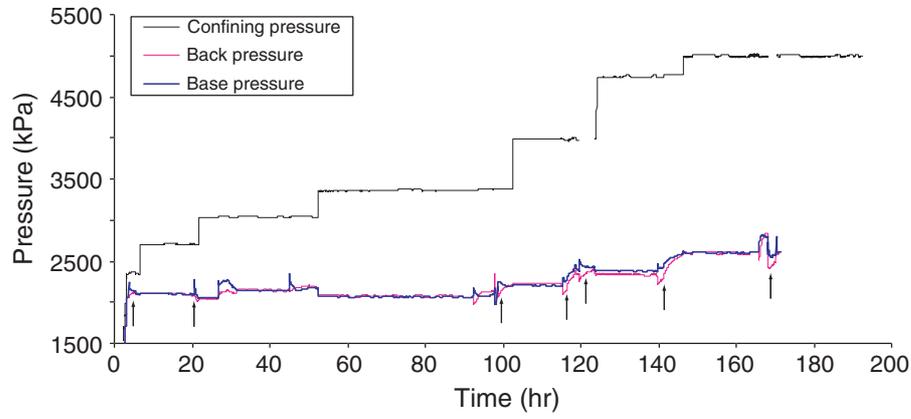

**Figure F5.** Sample 190-1173A-55X5, 135–150 cm (vertical), after the experiment.

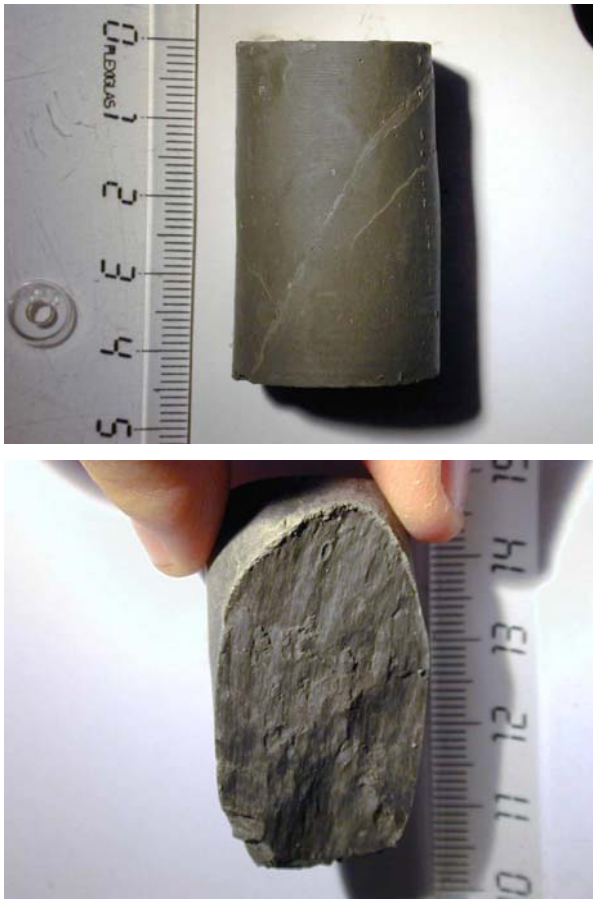

**Table T1.** Sample characteristics and estimated in situ overburden stress.

| Core, section, interval (cm) | Depth (mbsf) | Lithology | Porosity (%) | Length (mm) | Overburden stress (MPa) | Effective overburden stress (MPa) |
|---|---|---|---|---|---|---|
| 190-1174B-42R-1, 93–95 (v) | 535 | Clay and silt | 0.38 | 47.8 | 9.9 | 2.7–4.5 |
| 190-1173A-55X-5, 135–150 (v) | 520 | Clay and silt | 0.43 | 41 | 8.8 | 2.1–3.6 |
| 190-1173A-55X-5, 135–150 (h) | 520 | Clay and silt | 0.43 | 41.4 | 8.8 | 2.1–3.6 |

Note: v = vertical, h = horizontal.

**Table T2.** Permeability values and stress, strain conditions.

| Core, section | Permeability (m²) | Effective confining pressure (MPa) | Effective stress (MPa) | State | Deformation | Cumulative void ratio decrease |
|---|---|---|---|---|---|---|
| 190-1174B- | | | | | | |
| 42R-1 (v) | $3.6 \pm 0.1 \times 10^{-19}$ | 0.48 | 0.5 | Isotropic | | |
| 42R-1 (v) | $3.7 \pm 0.3 \times 10^{-19}$ | 0.5 | 0.5 | Isotropic | | |
| 42R-1 (v) | $1.2 \pm 0.2 \times 10^{-18}$‡ | 0.6 | 0.6 | Isotropic | | |
| 42R-1 (v) | $4.5 \pm 0.2 \times 10^{-19}$ | 0.5 | 0.5 | Isotropic | | ** |
| 42R-1 (v) | $4.1 \pm 0.15 \times 10^{-19}$‡ | 1.41 | 0.5 | Isotropic | | 0.013 |
| 42R-1 (v) | $1.6 \pm 0.1 \times 10^{-19}$ | 1.43 | 1.4 | Isotropic | | 0.013 |
| 42R-1 (v) | $3.7 \pm 0.3 \times 10^{-19}$ | 1.5 | 1.5 | Isotropic | | 0.015 |
| 190-1173A- | | | | | | |
| 55X-5 (v)* | $1.9 \pm 0.6 \times 10^{-19}$ | 0.12 | 0.15 | Isotropic | | |
| 55X-5 (v)* | $1.0 \pm 0.2 \times 10^{-19}$ | 0.21 | 0.2 | Isotropic | | |
| 55X-5 (v)* | $1.5 \pm 0.2 \times 10^{-18}$ | 0.21 | 1 | Deviatoric | Just fractured—0.015 | |
| 55X-5 (v)* | $1.3 \pm 0.2 \times 10^{-18}$ | 0.25 | 1.4 | Deviatoric | Just fractured—0.028 | |
| 55X-5 (v)† | $8.4 \pm 0.4 \times 10^{-19}$ | 0.22 | 0.35 | Isotropic | Previously fractured | ** |
| 55X-5 (v)† | $8.0 \pm 0.4 \times 10^{-19}$ | 0.56 | 0.55 | Isotropic | Previously fractured | 0.023 |
| 55X-5 (v)† | $6.4 \pm 0.4 \times 10^{-19}$ | 1.23 | 1.3 | Isotropic | Previously fractured | 0.044 |
| 55X-5 (v)† | $1.7 \pm 0.1 \times 10^{-19}$ | 1.57 | 1.6 | Isotropic | Previously fractured | 0.054 |
| 55X-5 (v)† | $1.4 \pm 0.2 \times 10^{-19}$ | 1.68 | 1.8 | Isotropic | Previously fractured | 0.054 |
| 55X-5 (v)† | $2.1 \pm 0.1 \times 10^{-19}$ | 2.41 | 2.2 | Isotropic | Previously fractured | 0.072 |
| 55X-5 (v)† | $1.9 \pm 0.1 \times 10^{-19}$ | 2.4 | 3.1 | Deviatoric | Previously fractured—0.0357 | 0.094 |
| 55X-5 (h) | $7.0 \pm 1.2 \times 10^{-19}$ | 1.1 | 1.1 | Isotropic | 0.0005 | 0.023 |
| 55X-5 (h) | $3.0 \pm 1.1 \times 10^{-19}$ | 2 | 2 | Isotropic | 0.0034 | 0.048 |
| 55X-5 (h) | $3.9 \pm 0.3 \times 10^{-19}$ | 1.94 | 2 | Isotropic | 0.0034 | 0.048 |
| 55X-5 (h) | $5.3 \pm 0.2 \times 10^{-19}$ | 2.41 | 2.4 | Isotropic | 0.0046 | 0.053 |
| 55X-5 (h) | $4.0 \pm 0.2 \times 10^{-19}$ | 2.43 | 2.4 | Isotropic | 0.0051 | 0.062 |
| 55X-5 (h) | $2.4 \pm 0.3 \times 10^{-19}$ | 2.43 | (5.0 to) 2.5 | Deviatoric | 0.015—after rupture | 0.072 |
| 55X-5 (h) | $3.8 \pm 0.2 \times 10^{-19}$ | 2.37 | 2.4 | Deviatoric | 0.015—after rupture | 0.072 |
| 55X-5 (h) | $4.1 \pm 0.3 \times 10^{-19}$ | 2.4 | 4.9 | Deviatoric | 0.0367—after rupture | 0.078 |

Notes: v = vertical, h = horizontal. Sample 190-1173A-55X-5, 135–150 cm (vertical), was subjected to two experiments: * = experiment 1, † = experiment 2. ‡ = steady-state flow measurements; ** = reference state for the calculation of the void ratio decrease. The cumulative void ratio for Sample 190-1173A-55X-5 (horizontal) is taken in comparison to the void ratio at 0.3 MPa effective pressure.

**Table T3.** Determination of friction coefficients.

| Core, section | Fracture angle (°) | (MPa) | | Axial strain range of the plateau (%) | q/p | (MPa) | | Friction coefficient |
|---|---|---|---|---|---|---|---|---|
| | | Axial effective stress | Confining effective stress | | | Normal stress | Shear stress | |
| 190-1173A- | | | | | | | | |
| 55X-5 (h) | 30 | 5.5 | 2.4 | 3.2–3.6 | 0.9 | 3.1 | 1.3 | 0.40 ± 0.15 |
| 55X-5 (v) | 35 | 4.9 | 2.4 | 3.1–3.6 | 0.77 | 3.2 | 1.2 | 0.37 ± 0.15 |
| 55X-5 (v) | 35 | 5.2 | 2.4 | 5.1–5.2 | 0.84 | 3.3 | 1.3 | 0.40 ± 0.15 |

Note: v = vertical, h = horizontal. q/p = deviatoric stress/mean effective stress ratio.